\documentclass[pdftex,twocolumn,epjc3]{svjour3}          

\newcommand\aem{\alpha_{\rm em}}

\RequirePackage[T1]{fontenc}

\smartqed  

\RequirePackage{graphicx}
\RequirePackage{mathptmx}      
\RequirePackage{flushend}
\RequirePackage[numbers,sort&compress]{natbib}
\RequirePackage[colorlinks,citecolor=blue,urlcolor=blue,linkcolor=blue]{hyperref}

\journalname{Eur. Phys. J. C}

\begin{document}

\title{Neutral current Drell--Yan with combined QCD and electroweak corrections 
in the POWHEG BOX}

\author{Luca Barz\`e\thanksref{e1,addr1}
\and
Guido Montagna\thanksref{e2,addr2}
 \and
Paolo Nason\thanksref{e3,addr3}
\and
Oreste Nicrosini\thanksref{e4,addr4}
\and\\
Fulvio Piccinini\thanksref{e5,addr4}
\and
Alessandro Vicini\thanksref{e6,addr5,addr1}
}

\thankstext{e1}{e-mail: luca.barze@cern.ch}
\thankstext{e2}{e-mail: guido.montagna@pv.infn.it}
\thankstext{e3}{e-mail: paolo.nason@mib.infn.it}
\thankstext{e4}{e-mail: oreste.nicrosini@pv.infn.it}
\thankstext{e5}{e-mail: fulvio.piccinini@pv.infn.it}
\thankstext{e6}{e-mail: alessandro.vicini@mi.infn.it}

\institute{PH-TH Department, CERN, CH 1211, Geneva 23, Switzerland\label{addr1}
\and
Dipartimento di Fisica, Universit\`a di Pavia and INFN, Sezione di Pavia, Via A. Bassi 6, 27100 Pavia, Italy\label{addr2}
 \and
INFN, Sezione di Milano Bicocca and Dipartimento di Fisica, Universit\`a di Milano Bicocca, 20133 Milano, Italy\label{addr3}
\and
INFN, Sezione di Pavia,  Via A. Bassi 6, 27100 Pavia, Italy\label{addr4}
\and
Dipartimento di Fisica, Universit\`a di Milano and INFN, Sezione di Milano, Via Celoria 16, 20133 Milano, Italy\label{addr5}
}

\date{Received: date / Accepted: date}

\maketitle

\begin{abstract}
Following recent work on the combination of electroweak and strong radiative corrections 
to single $W$-boson hadro\-production in the POWHEG BOX framework, we generalize the 
above treatment to cover the neutral current Drell-Yan process. According to the POWHEG method, 
we combine both the next-to-leading order (NLO) electroweak and QED multiple photon corrections 
with the native NLO and Parton Shower QCD contributions. We show comparisons with the predictions of 
the electroweak generator HORACE, to validate the reliability and accuracy of the approach.
We also present phenomenological results obtained with the new tool for physics studies at the LHC.
\end{abstract}

\section{Introduction}
\label{section:introduction}

The majority of the searches and studies at the proton-proton ($p p$) collider LHC at CERN, including the 
analyses related to the Higgs boson, is based on an intensive use of Parton Shower 
(PS) generators matched 
with fixed-order perturbative calculations. The success of these general computational frameworks, like 
MC@NLO~\cite{Frixione:2002ik} and POWHEG~\cite{Nason:2004rx,Frixione:2007vw}, rests on their
capability to provide reliable predictions for both inclusive cross sections and distributions of a large 
variety of signatures in the presence of arbitrary event selection conditions.
The increased theoretical accuracy with respect to traditional PS generators has promoted
the wide use of these frameworks on the experimental side. Generally 
speaking, the  processes available in such tools include finite-order contributions at the
next-to-leading order (NLO) accuracy in QCD \cite{Nason:2012pr}.
Only recently the implementation of electroweak (EW) 
contributions 
in matched PS generators has been addressed in the literature~\cite{Barze:2012tt,Bernaciak:2012hj}
in the process of single $W$-boson hadroproduction, according to the matching approach given by the
POWHEG method. High precision in the $W$ hadroproduction process is required 
in particular by the need to reduce theoretical uncertainties in $W$ mass measurements. 
The two implementations describ\-ed in Refs. \cite{Barze:2012tt,Bernaciak:2012hj} are both
available as subprocesses in the repository\footnote{See http://powhegbox.mib.infn.it for 
an updated list of all available processes.} of the general computer  framework
POWHEG BOX~\cite{Alioli:2010xd}. 

For many physics studies at the LHC, also the process of  lepton pair production in hadronic collisions, known as 
neutral current Drell-Yan (NC DY), requires
high theoretical accuracy. The state of the art of fixed-order calculations is encoded 
in the next-to-next-to-leading order (NNLO) QCD programs DYNNLO~\cite{Catani:2009sm} 
and FEWZ~\cite{Melnikov:2006kv,Gavin:2010az}, 
in the EW codes HORACE~\cite{CarloniCalame:2006zq,CarloniCalame:2007cd} and 
ZGRAD/ZGRAD2~\cite{Baur:1997wa,Baur:2001ze}, 
while the SANC framework~\cite{Arbuzov:2007db,Andonov:2008ga,Bardin:2012jk,Bondarenko:2013nu} 
and the RADY code~\cite{Dittmaier:2009cr} allow
to evaluate both QCD and EW NLO corrections. Recently, NNLO QCD corrections were combined 
with NLO EW contributions 
in FEWZ~\cite{Li:2012wn}.

All these theoretical efforts were made because, given the high-precision measurement of 
the $Z$-boson mass at LEP, 
the NC process is very helpful, if not unavoidable, for detector calibration purposes at LHC and 
is a standard candle that can be used to constrain the Parton Distribution Functions (PDFs).
Moreover, the transverse momentum distribution of the $Z$ boson can be accurately measured, and used to tune
non-perturbative parameters in the generators. This measurement indirectly constrains the transverse
momentum distribution of the $W$ boson, with an accuracy that strictly depends upon the accuracy 
of the production model.
 The NC DY process also allows to perform tests of the SM at the loop level and to measure EW parameters, such as the
 weak mixing angle, in a hadronic environment. In the high tail of the transverse momentum and 
 invariant mass distributions of the produced leptons, the NC DY is one of the main irreducible backgrounds  
 to the searches for new particles at the LHC. 
 
 These physics motivations require the simultaneous control 
 of all the relevant QCD and EW higher-order contributions to the NC DY process. Following the 
 recent implementation of EW corrections to $W$ production in the POWHEG BOX as 
 described in Ref. \cite{Barze:2012tt}, here we present the analogous treatment for the NC DY process.
In particular, the phenomenological importance of combining QCD and EW corrections to
dilepton hadroproduction is illustrated through the analysis of various numerical results. Let us note that
our approach is complementary to the combination of NLO EW corrections to NC DY with 
fixed-order QCD at NLO and NNLO accuracy as recently realized in the SANC 
framework \cite{Bardin:2012jk,Bondarenko:2013nu} and
in the FEWZ code~\cite{Li:2012wn}, respectively\footnote{Further studies on
combining QCD and EW corrections to $W/Z$ hadroproduction are described in 
Refs.~\cite{Cao:2004yy,Adam:2008ge,Adam:2008pc,Balossini:2008cs,Balossini:2009sa,Richardson:2010gz,Yost:2012az}}.  
However, unlike the above additive combinations, mixed EW-QCD corrections, as well as QCD and 
QED shower effects, are taken into account in our realization. Than\-ks to it, it is therefore possible to obtain 
reliable predictions in the presence of combined QCD and EW effects also for those
observables, like the $Z$ and lepton transverse momentum, 
which are particularly sensitive 
to logarithmically enhanced higher-order QCD contributions and are not realistically accessible to the tools including
fixed-order corrections only.

The paper is organized as follows. In Section \ref{section:calculation} we describe the
modifications applied to the POWHEG BOX for the inclusion of the NLO EW and higher-order 
QED corrections in the NC DY channel. In Section \ref{section:pheno} we present and discuss
several numerical results obtained with the new tool at LHC energies, both as a cross-check of the EW
corrections at NLO accuracy and about the interplay of QCD, QED and weak corrections. Conclusions
are given in Section \ref{section:conclusions}.

\section{Details of the calculation}
\label{section:calculation}

POWHEG (POsitive Weight Hardest Emission Generator) is a method conceived for embedding NLO QCD computations 
into PS simulations. Here we just review the basic ingredients and ideas of the method, paying particular 
attention to the components that we generalized for the inclusion of the EW and QED corrections. 
For more details the reader is referred to the original literature~\cite{Nason:2004rx,Frixione:2007vw,Alioli:2010xd}.

\subsection{The POWHEG method}

In the POWHEG formalism, the generation of the hardest emission is performed first,
using full NLO accuracy, and a shower Monte Carlo is used to generate subsequent radiation.
Therefore, the building blocks of the method are those typical of a NLO calculation and 
of PS generators. 

At NLO accuracy in QCD, the necessary ingredients are given by the process-dependent 
virtual corrections $V_{QCD}$ and real radiation matrix element(s) $R_{QCD}$. 
To ensure cancellation of the initial-state collinear singularities in hadronic collisions, further 
components of the calculation are two factorization counterterms, one for each of the incoming partons ($\oplus, \ominus$), 
which we denote, following the POWHEG notation, as $G_{\oplus,\ominus}^{QCD}$ (the collinear remnants). Soft and collinear 
divergences coming from real radiation are treated in POWHEG 
using the FKS subtraction formalism~\cite{Frixione:1995ms}.  The method requires that the real radiation contribution is 
separated into a sum of terms $R^{\alpha}$ (where $\alpha$ labels all the singular
regions of the real amplitude) such that each $R^{\alpha}$ is singular
only in the  $\alpha$ region. The formalism is completed by the introduction of 
a set of functions $C_{QCD}^{(\alpha)}$ which play the role of real counterterms.


With all these ingredients at hand, the calculation of an inclusive 
cross section with NLO QCD accuracy proceeds in the POWHEG BOX according to a high degree of automation. 
First the algorithm identifies all the singular regions and maps the real radiation matrix element over the singular 
configurations. Afterwards, it performs the subtraction procedure and computes the collinear remnants. As shown 
in the following, this procedure can be effectively generalized to the inclusion of EW contributions, provided 
the appropriate ingredients and modifications are supplied. 

In the POWHEG method, the matching of the NLO computation with the PS is achieved in terms of a modified 
Sudakov form factor, which contains the NLO real radiation matrix element and 
is equal to the product of the Sudakov form factors for each singular region. The 
generation of the event with the hardest radiation, which is a fundamental aspect of the approach, is also 
handled automatically by the POWHEG BOX framework and can be adapted to cope with the 
radiation of colorless partons, as discussed in the following.

\subsection{Electroweak contributions: calculation and inclusion in the POWHEG BOX}

We treated the NC DY process according to the same recipe adopted in Ref. \cite{Barze:2012tt} 
for the implementation of the EW contributions 
to single $W$ production in the POWHEG BOX. 

The most intricate and lengthy step is the calculation of the one-loop EW corrections $V_{EW}$, since 
it involves the treatment of unstable particles in addition to the cancellation of mass singularities.
We accomplished this task realizing a library for the computation of $V_{EW}$, 
based upon the one-loop structure for the virtual photonic and weak corrections given in Ref. \cite{Dittmaier:2009cr}. Unlike
Ref. \cite{Dittmaier:2009cr}, where mass regularization is used to treat soft and collinear singularities, we 
employed a mixed scheme in order to comply with the use of dimensional regularization of POWHEG.
Precisely, we used dimensional regularization to treat the singularities associated to the QCD partons and
the photon, but we kept the mass of the leptons finite, since this mass has a well-defined physical meaning, 
being the true regulator of the QED mass singularities. For the evaluation of the
scalar integrals, we resorted in particular to various results of Ref. \cite{Dittmaier:2003bc}
(for the computation of the three-point functions) and Ref. \cite{Denner:2010tr}
(for the computation of the four-point functions). 

\begin{figure}[h] 
\vspace{-2.5cm}
\hspace{-1.5cm}\includegraphics[height=16.cm]{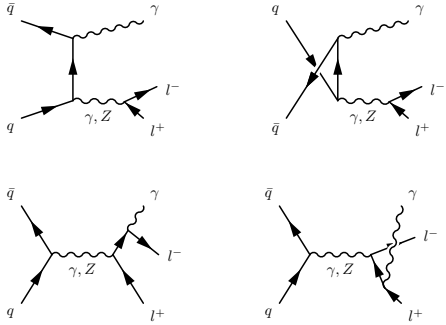}
\vspace{-9.5cm}
\caption{The photon bremsstrahlung diagrams of the NC DY process.}
\label{feynmanb}
\end{figure}

The singularities associated to the unstable nature of the weak bosons circulating in the 
loops and given by logarithms of the form $\log (s - M^2_V + i \epsilon)$ were treated according to two
different schemes, which both respect gauge invariance. In the first 
approach (factorization scheme) \cite{Dittmaier:2009cr,Dittmaier:2001ay},
the logarithms that are singular at the resonance (and would be cured by a Dyson resummation of the 
self-energy contributions inside the loop diagrams) are treated with the substitution 
 $\log (s - M^2_V + i \epsilon) \to  \log (s - M^2_V + i \Gamma_V M_V), V = W,Z$. In the second approach 
 (complex mass scheme)~\cite{Denner:2005fg,Denner:2006ic}, the squared vector boson 
 masses are taken as complex quantities $\mu_{W,Z}^2 = M_{W,Z}^2 - i \Gamma_{W,Z} M_{W,Z}$ 
 in the LO and NLO calculation, leading
 to complex couplings as well. Both schemes were implemented in the POWHEG BOX and 
 the differences between the two procedures for treating the resonance are at the per mille level or 
 below for arbitrary partonic centre of mass (c.m.) energies $\hat{s}$~\cite{Dittmaier:2009cr}.

To complete the NLO EW calculation, we computed the real photon matrix element
$R_{EW}$ relative to the bremss\-tra\-hlung processes $q \bar{q} \to \gamma, Z \to \to l^+ l^- \gamma$ 
shown in Fig.~\ref{feynmanb}. 
We included finite lepton mass effects in the calculation and resorted to the algorithm developed 
and described in detail in Ref.~\cite{Barze:2012tt} 
for the treatment of the radiation phase space with massive fermions. The search for
the singular regions was extended to handle the divergences associated to photon 
radiation. The original algorithm applied to the initial-state radiation (ISR) of QCD partons 
was generalized to treat the mechanisms of QED ISR and final-state radiation (FSR). As
the large logarithms associated to photon ISR have to be reabsorbed in the PDFs, in analogy
to QCD, the dominant QED contribution is due to FSR, which contains logarithms of the
form $\alpha_{\rm em} \log({\hat{s}}/m_l^2)$ for the typical non-fully inclusive and non-infrared safe 
conditions used in the event selection. As demonstrated in Ref.~\cite{Barze:2012tt},
the new phase space treatment including massive leptons is a key ingredient to 
control the above logarithmic contributions. To ensure cancellation of all the mass
singularities in terms of the new phase space, we also recalculated the QCD matrix element $R_{QCD}$ including 
finite lepton mass 
contributions in the calculation of both the gluon emission $q \bar{q} \to \gamma, Z \to \to l^+ l^- g$
and gluon-induced $g q/{\bar q} \to l^+ l^-  q/{\bar q}$ processes. 

To summarize, the implementation of EW corrections in the POWHEG BOX was 
realized through the following generalization of the native QCD elements
\begin{eqnarray}
&& V_{QCD} \to V_{QCD} + V_{EW} \nonumber\\
&& R_{QCD} \to R_{QCD} + R_{EW} \nonumber\\
&& C_{QCD}^{(\alpha)} \to C_{QCD}^{(\alpha)} + C_{EW}^{(\alpha)} \nonumber\\
&& G_{\oplus,\ominus}^{QCD} \to G_{\oplus,\ominus}^{QCD} + G_{\oplus,\ominus}^{EW}
\end{eqnarray}
and improving the algorithm for the identification of the singular regions associated to real photon radiation.

For the cross section calculation, we implemented three input parameter schemes: the $G_\mu$, the on-shell $\aem (0)$ and 
the $\aem(M_Z)$ sch\-eme. 
In the first scheme, the primary input parameters are given by the muon decay constant and the weak boson mas\-ses; 
in the second one, $G_\mu$ is replaced by the fine structure constant at zero momentum transfer. In the latter scheme, 
the non-perturbative hadronic contribution to the running of $\aem$ is treated in terms of effective quark masses
reproducing the correct value of $\aem(Q^2)$ at high energies, i.e. $Q^2 = M_Z^2$. In the third scheme, 
the input parameters are $\aem$ at the scale $Q^2 = M_Z^2$, $M_Z$ and $M_W$. 
The widths of the weak bosons are kept fixed everywhere. 

As a last step, we matched the NLO QCD and EW corrections with PS contributions. We considered 
both QCD and QED parton cascade. According to POWHEG, the generation of the hardest-radiation 
event is performed in terms of the exact real radiation matrix element and of a modified Sudakov 
form factor including it. Once the configuration with the hardest (transverse momentum $p_\perp$) 
emission has been generated, 
the subsequent radiation processes handled by the PS take place at lower $p_\perp$, applying a veto technique. 
The above algorithm was generalized to include  photon radiation in the POWHEG BOX. We made use 
of both $R_{QCD}$ and $R_{EW}$ and implemented a Sudakov form factor equal to the product of the 
form factors for ISR gluon and photon radiation,
as well as FSR photon radiation. In the native QCD construction, the method requires a lower cut-off on the 
transverse momentum, 
in order to avoid to reach unphysical values of the strong coupling constant and of the PDFs. 
Also this requirement was generalized, following the procedure applied to $W$
production in Ref. \cite{Barze:2012tt}. Precisely, the lower $p_\perp$ cut-off was set equal to a typical hadronic
scale for gluon or photon radiation from quarks, while it was taken as the mass of the lepton
for QED radiation off the leptons. The package PHOTOS \cite{Golonka:2005pn}\footnote{We use PHOTOS 
with leading log kernels, i.e. 
without process-dependent matrix element corrections already included in our approach in the 
exact NLO calculation.} 
is used to handle multiple photon emission, 
with enforced $p_\perp$-ordered radiation.

\begin{figure}[h] 
\vspace{-2.5cm}
\hspace{-1.5cm}
\includegraphics[height=16.cm]{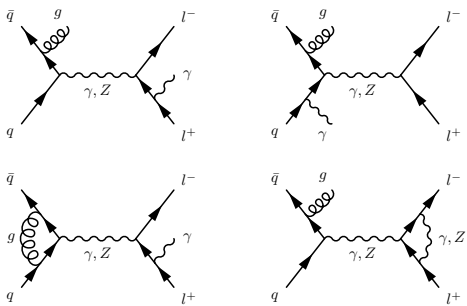}
\vspace{-9.5cm}
\caption{Examples of mixed QCD$\otimes$EW contributions included in the POWHEG BOX assuming factorization.}
\label{feynman-mixed}
\end{figure}

Concerning the treatment of multiple QCD and QED radiation in our approach, a few comments are in order here. 
The choice of using two separate showers for QCD and QED radiation is
motivated by the need of providing an accurate modeling of the mechanism
of QED FSR (beyond one-photon emission) as ensured by PHOTOS. On the other hand,
the use of two separate generators suffers of the drawback of neglecting
the contribution of the emission of a second IS photon after radiation of
a parton or a photon, being the latter included in our approach using exact
real radiation matrix elements. This approximation amounts to neglecting terms of 
$O(\alpha_s \aem^n), n \geq 1$ and of $O(\aem^n), n \geq 2$ at the level of ISR.  
However, this prescription should be sufficiently accurate because FS photon radiation largely dominates over IS 
QED radiation and mixed $O(\alpha_s^n \aem^m)$ due to the interplay between IS QCD and FS QED 
multiple radiation
are included in our formulation. The accuracy of this approach could be validated through comparisons with 
the results of a single shower handling QCD and QED radiation simultaneously, as available e.g. in PYTHIA.


\begin{figure}[h] 
\vspace{-2cm}
\hspace{-2.5cm}
\includegraphics[height=17.cm]{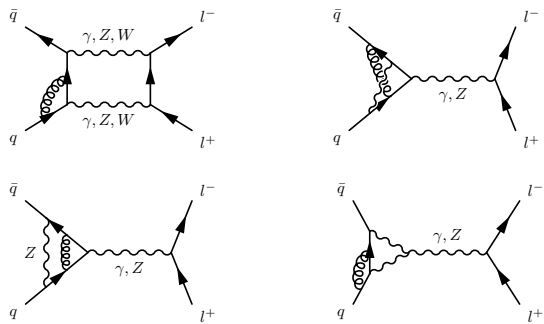}
\vspace{-10.cm}
\caption{Examples of Feynman diagrams contributing to $O(\alpha_{\rm em} \alpha_s)$ corrections 
beyond the LL accuracy not accounted for in our approach.}
\label{feynman-mixed-new}
\end{figure}

\begin{table*}
\caption{The input parameters used in the numerical simulations.}
\label{input}
\begin{tabular*}{\textwidth}{@{\extracolsep{\fill}}lll@{}}
\hline
\hline
$ \aem (0) =1/137.03599911$ & $M_Z = 91.1876$~GeV  &
 $M_W = 80.37399$~GeV \\ 
$\Gamma_Z = 2.4924$~GeV & $\sin^2\theta_W = 1 - M_W^2/M_Z^2$& $M_{\rm Higgs} = 125$~GeV \\
$\Gamma_W = 2.141$~GeV & $\alpha_s (M_Z) = 0.1205$ & \\
$m_e=510.99892$~KeV & $m_{\mu}=105.658369$~MeV & $m_{\tau}=1.77699$~GeV \\
$m_u = 0.06983$~GeV & $m_c = 1.2$~GeV & $m_t = 174$~GeV \\
$m_d = 0.06984$~GeV & $m_s = 150$~MeV & $m_b = 4.6$~GeV \\
\hline
\end{tabular*}
\end{table*}

In conclusion, let us note that the inclusion of all the theoretical elements as described above
allows to obtain predictions for dilepton production in hadronic collisions taking into account the 
interplay between QCD and EW contributions in a single computational framework. 
In particular, the matching of the complete NLO SM corrections with 
the showers of both colored particles and photons enables to perform simulations of the differential cross sections
in the presence of mixed QCD$\otimes$EW effects, in addition to the contribution of 
higher-order QCD and QED corrections. In the absence of a complete calculation of 
$O(\alpha_{\rm em} \alpha_s)$ corrections to DY processes\footnote{Partial results exist in the 
literature and are given by the one-loop EW corrections to $W/Z$+jet production at finite transverse 
momentum~\cite{Kuhn:2007qc,Kuhn:2007cv,Hollik:2007sq,Denner:2009gj}, NLO QCD corrections to 
the $W/Z+\gamma$ process~\cite{Dixon:1998py,DeFlorian:2000sg,Campbell:2011bn} and 
two-loop virtual $O(\alpha_{\rm em} \alpha_s)$ corrections to the NC DY \cite{Kilgore:2011pa}. However, the 
complete and non-trivial combination of all the above substructures is still unavailable.}, the tool enables to obtain, 
using factorization, leading logarithmic (LL) predictions 
for mixed QCD$\otimes$EW contributions, as schematically 
shown in Fig.~\ref{feynman-mixed}. This approach can be expected to
provide accurate results when soft and collinear radiation dominates and
factorization arguments apply. Therefore, our QCD$\otimes$EW combination can be considered
strictly reliable in the LL approximation. An exact control of the coefficients of 
subleading corrections and of constant terms at 
$O(\alpha_{\rm em} \alpha_s)$ would require a complete calculation
at the two-loop level, which is unavailable yet. 
Examples of contributions not included in our calculation 
come from the first two diagrams of Fig.~\ref{feynman-mixed}
when both QCD and QED emission is related to hard 
gluon and photon radiation at large angles. Other examples of 
missing corrections are shown in Fig.~\ref{feynman-mixed-new}. 
As they describe virtual insertions involving off-shell 
fermion lines, they contribute to mixed corrections beyond the LL accuracy.

\section{Phenomenological results}
\label{section:pheno}

In this Section, we show the results of a variety of tests performed to validate the new theoretical and
computational ingredients implemented in the POWHEG BOX. We first present comparisons against
the benchmark predictions of the HORACE generator \cite{CarloniCalame:2006zq,CarloniCalame:2007cd} 
at NLO EW accuracy\footnote{Note that HORACE implements
completely independent EW form factors and real photon matrix elements, computed in
the mass regularization scheme.}.
Then we address a phenomenological 
analysis of the interplay between QCD and EW corrections of interest for physics studies at the LHC.

\subsection{Input parameters and event selection}

We considered the lepton pair hadroproduction process $p p \to \gamma,Z \to \mu^+ \mu^- + (X)$, 
at the c.m. energy $\sqrt{s} = 14$~TeV. 
We used the MRST2004QED set of PDFs \cite{Martin:2004dh} with renormalization and 
(QCD and QED) factorization scale $\mu_R = \mu_F = M_Z$
for the comparison with HORACE and the QCD$\otimes$EW simulations around the 
$Z$ resonance, and $\mu_R = \mu_F = M_{l^+ l^-}$ (the lepton pair 
invariant mass) for the numerical results on the combination of QCD and EW corrections well above the peak.
For both the final-state leptons, we applied the following cuts on the transverse momentum and pseudorapidity:
\begin{eqnarray}
p_\perp^{\mu^{\pm}} > 20 \, \, {\rm GeV} \, , \qquad \qquad |\eta_{\mu^{\pm}}| < 2.5 \, ,
\label{eq:cuts}
\end{eqnarray}
which approximately model the acceptance of the ATLAS and CMS detectors at the LHC. 
In addition we also applied 
a cut on the invariant mass of the lepton pair of $M_{l^+ l^-}$ > 50~GeV.
We considered, for simplicity, ``bare" leptons, {\it i.e.} nearby photons are not recombined with the leptons. 
The results have been 
obtained in the $\aem (0)$ scheme, using the input parameters listed in 
Table~\ref{input}\footnote{The CKM mixing matrix is 
set to the identity matrix in the calculation of EW loop corrections.}, 
which coincide with those adopted in the tuned comparisons of EW predictions for $Z$ boson
observables described in Ref.~\cite{Buttar:2008jx}. The overall choice of input parameters, PDF set and 
acceptance cuts allowed us to check first that the NLO EW predictions of both HORACE and the 
new POWHEG version perfectly agree with the benchmark results for the integrated cross sections
given in Ref.~\cite{Buttar:2008jx}.  For the QCD results and those about 
the combination of QCD and EW contributions we use $\alpha_s (M_Z) = 0.1205$ with NLO evolution.


\subsection{Comparisons with HORACE}

\begin{figure} 
\begin{center}
\includegraphics[height=6.cm]{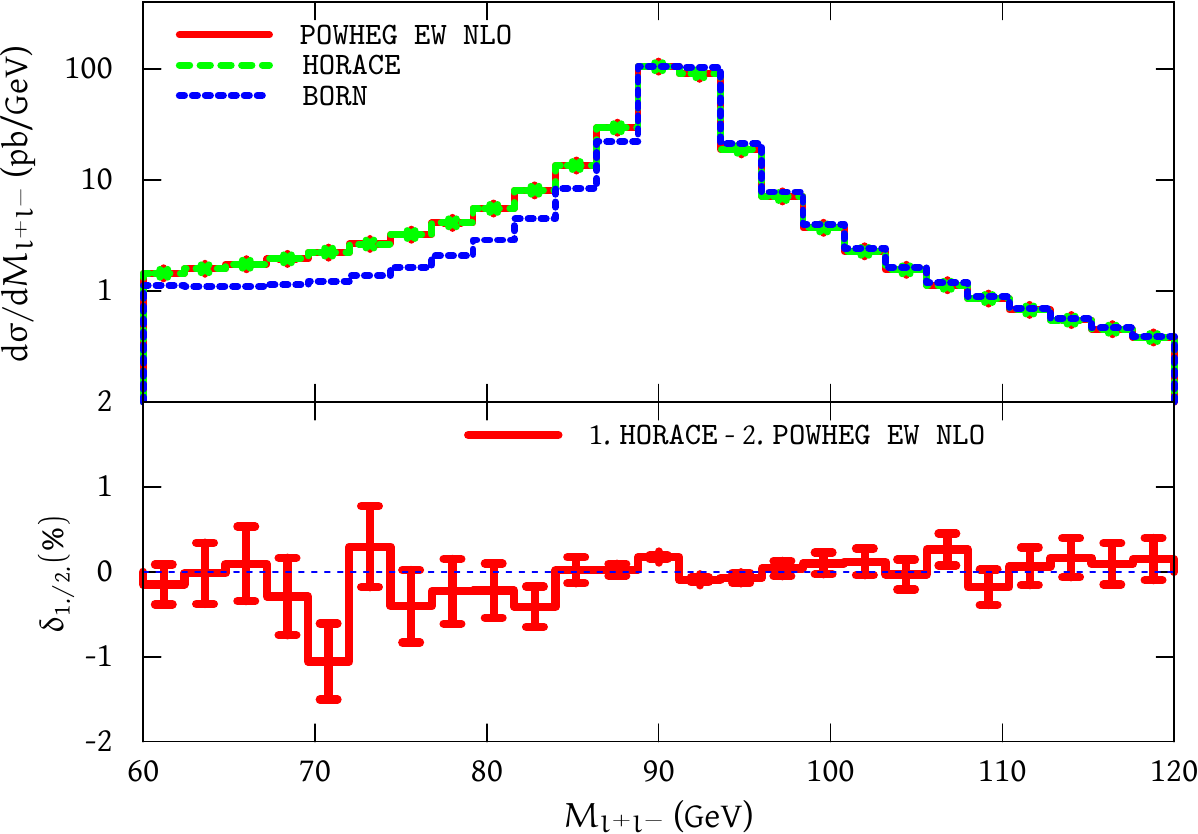}
\end{center}
\caption{Upper panel: lepton-pair invariant mass distribution according to POWHEG BOX and HORACE
at NLO EW accuracy, in the resonance region. Lower panel: relative deviations, in percent, 
between the predictions of the two generators.}
\label{horace1}
\end{figure}

\begin{figure} 
\begin{center}
\includegraphics[height=6.cm]{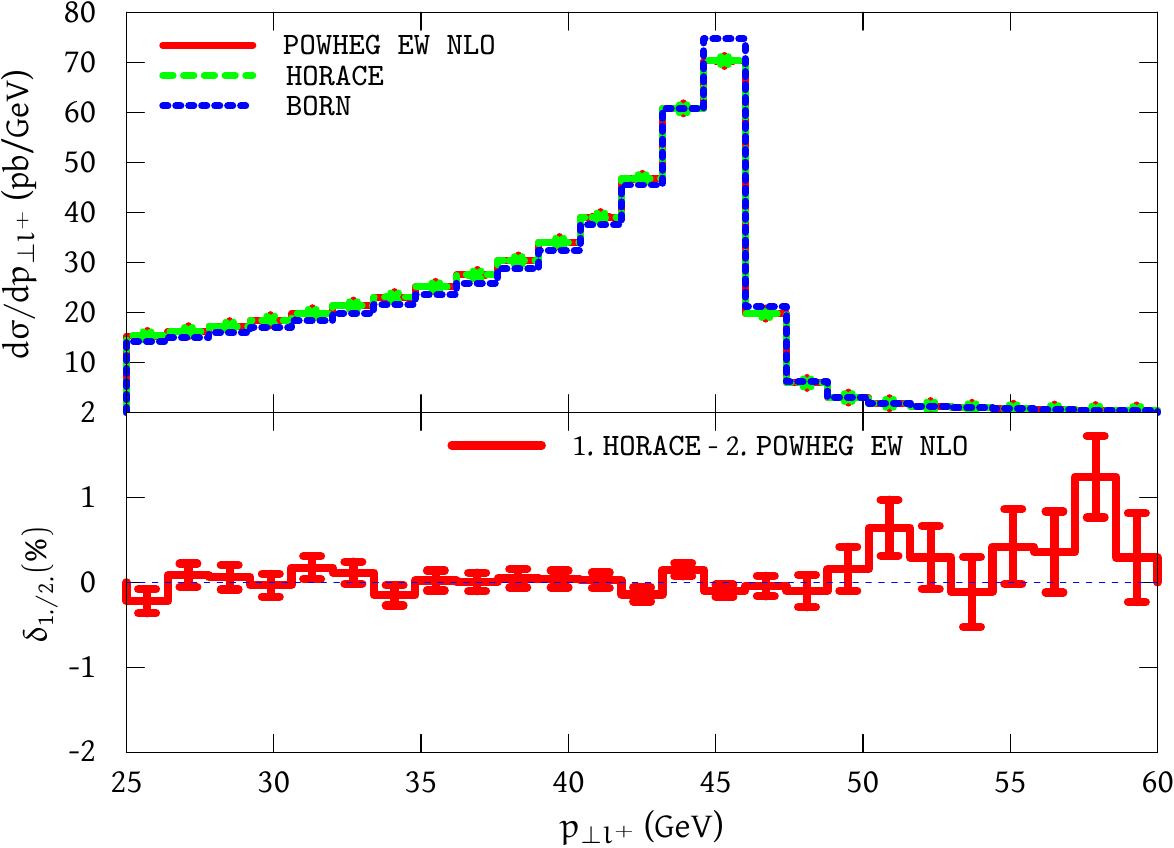}
\end{center}
\caption{The same as Fig. \ref{horace1} for the lepton transverse momentum distribution.}
\label{horace2}
\end{figure}

\begin{figure} 
\begin{center}
\includegraphics[height=6.cm]{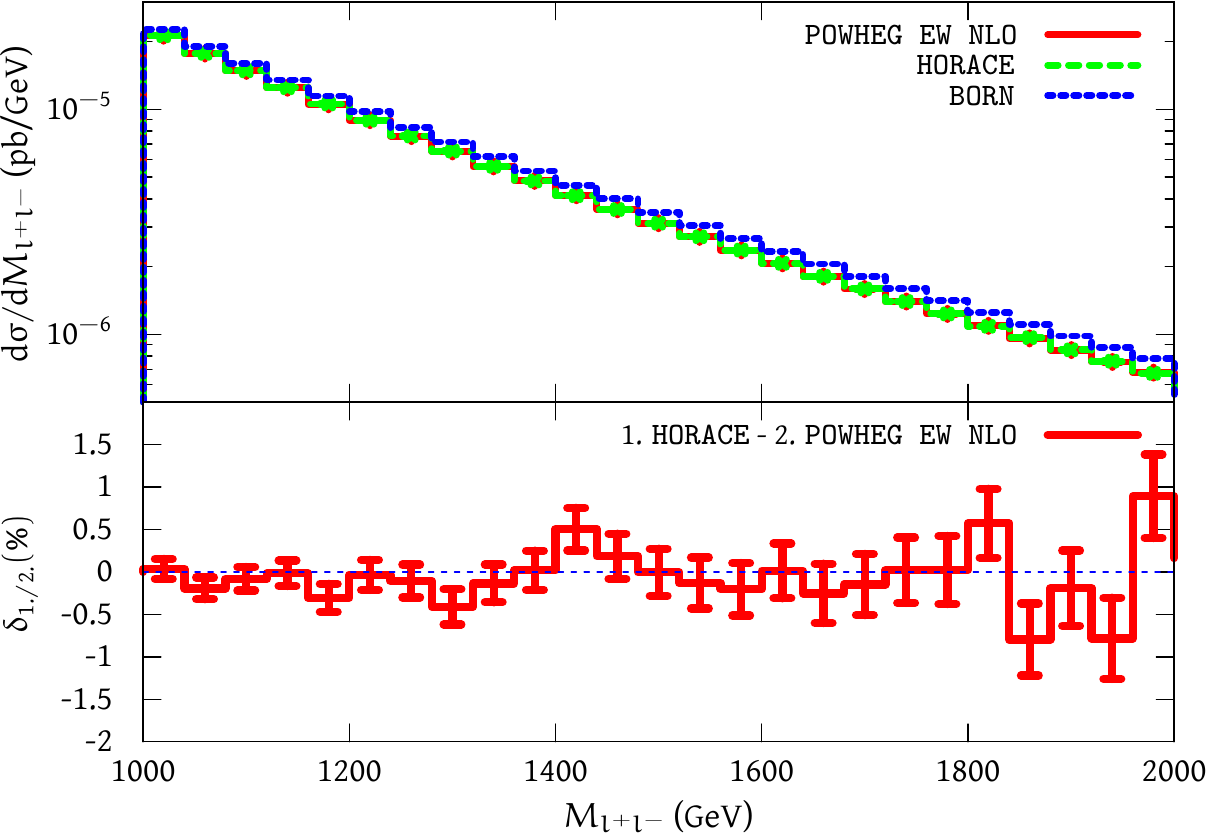}
\end{center}
\caption{The same as Fig. \ref{horace1} in the region $M_{l^+ l^-}$ > 1~TeV.}
\label{horace3}
\end{figure}

\begin{figure} 
\begin{center}
\includegraphics[height=6.cm]{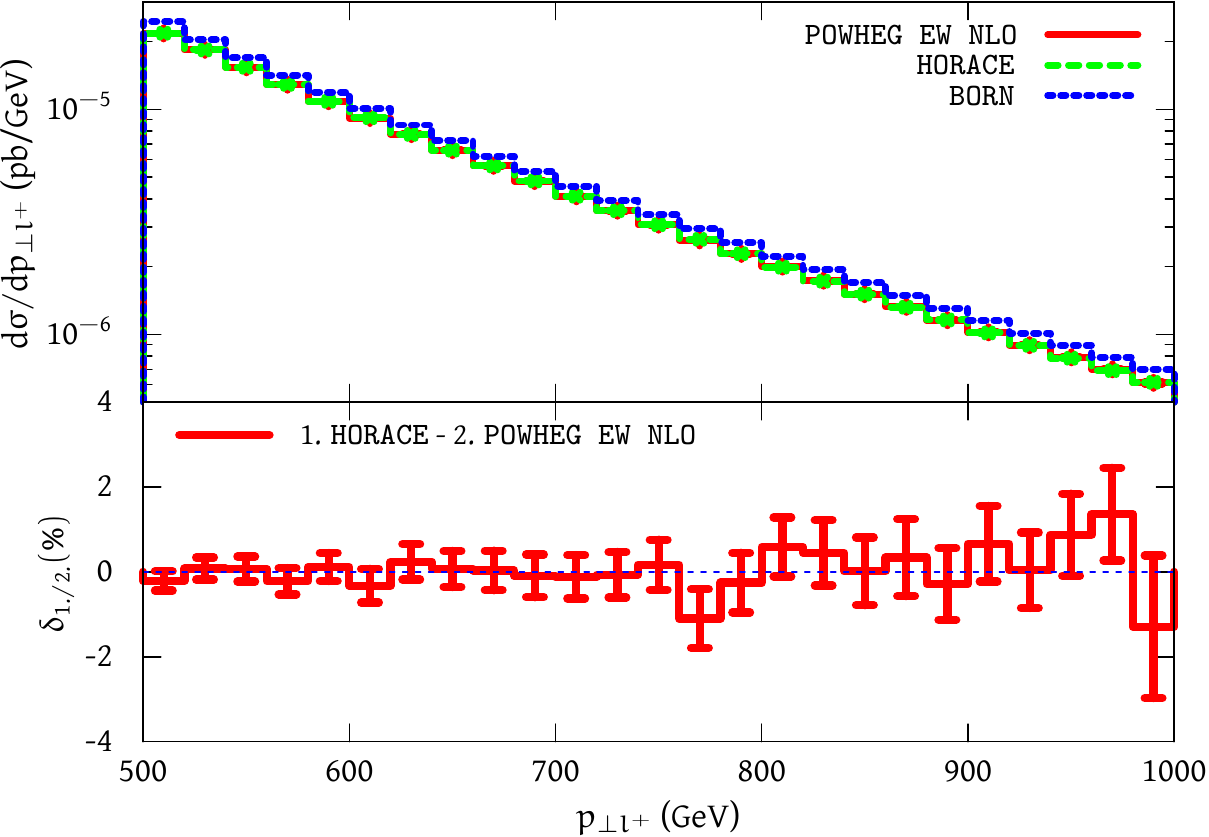}
\end{center}
\caption{The same as Fig. \ref{horace3} for the lepton transverse momentum distribution.}
\label{horace4}
\end{figure}

The results of the validation of the pure NLO 
EW corrections in the POWHEG BOX framework 
are shown in Figs.~\ref{horace1} through \ref{horace4} 
by comparisons with HORACE simulations.  The invariant mass
region around the $Z$ resonance and well 
above it ($M_{l^+ l^-}$ > 1~TeV) are considered 
 in Fig.~\ref{horace1} - Fig. \ref{horace2} and Fig.~\ref{horace3} - Fig.~\ref{horace4}, respectively. 
 For these comparisons we took 
 the limit $\alpha_s \to 0$ numerically in POWHEG. In the upper panels, we show the 
NLO predictions of the two codes for 
the invariant mass and $p_\perp^l$ distributions, together
with the reference Born results. The lower panels show the relative difference 
(in percent) between the two programs, where the error bars correspond to 1$\sigma$ 
numerical uncertainties. 

We observe that the results of the two programs are in good agreement, both in the peak
region and in the high tails. This means that all the EW ingredients have 
been correctly included in POWHEG. In particular, the agreement in the $Z$ resonance 
region is a test of the correct calculation of the NLO virtual and real contributions, as well 
as of the generalized subtraction procedure and treatment of collinear photon singularities. 
Indeed, the EW corrections in the peak region are known to be dominated 
by the mechanism of final state photon radiation, which receives contributions from all
the above components. On the other hand, the NLO EW corrections to the 
tails of the distributions give rise to large negative contributions due to Sudakov-like 
logarithms of the form $\aem \log(\hat{s}/M_V^2)$. The latter come from the exchange 
of weak gauge bosons in the loops, 
which therefore display in POWHEG the expected behavior at high energies\footnote{Note that we 
do not take into account in our calculation the contribution due to real radiation of massive gauge bosons, which is 
known~\cite{Baur:2006sn,Bell:2010gi} to partially compensate the negative virtual Sudakov-like corrections.}.

In summary, the implementation in the POWHEG BOX of all the NLO EW corrections to the NC DY 
can be considered fully under control.

\subsection{Combined effect of QCD and EW corrections}

\begin{figure} 
\begin{center}
\includegraphics[height=6.cm]{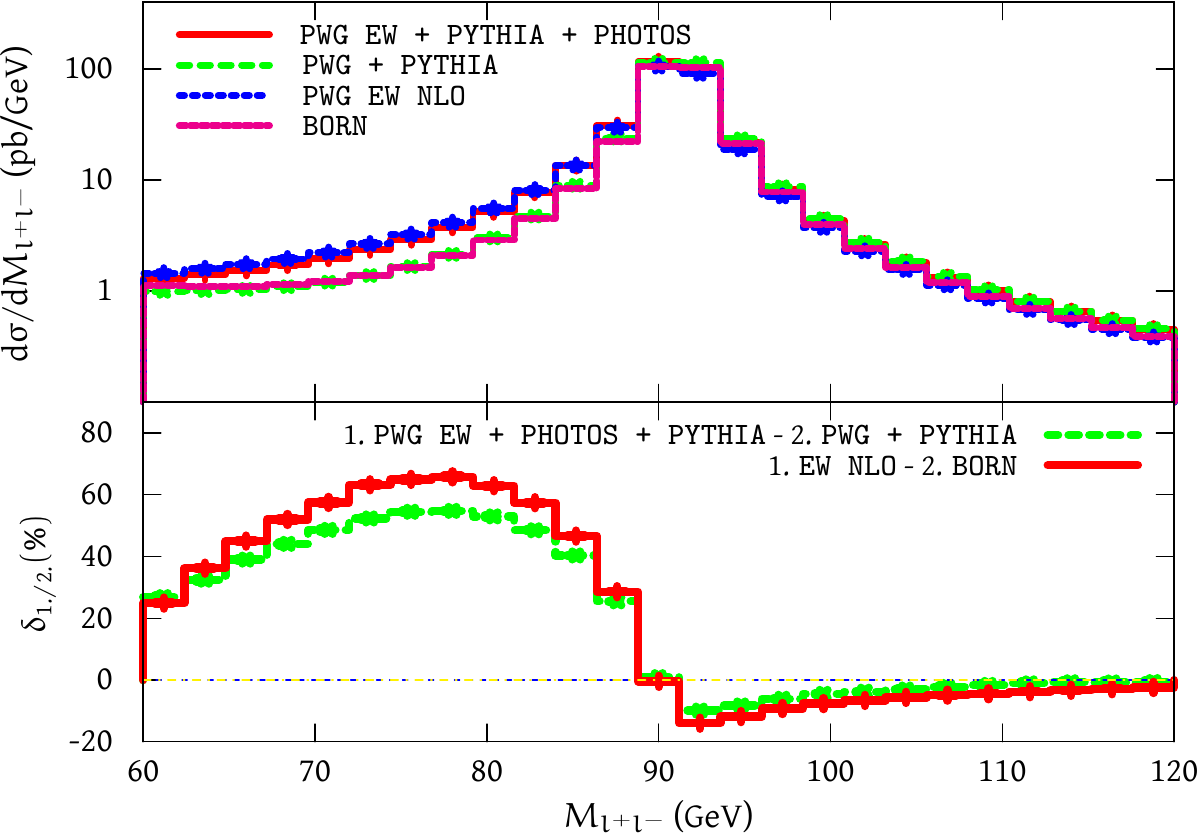}
\end{center}
\caption{Upper panel: lepton-pair invariant mass distribution around the resonance according 
to the full QCD$\otimes$EW predictions of the POWHEG BOX (PWG EW + PYTHIA+PHOTOS), 
the standard QCD POWHEG BOX (PWG + PYTHIA), the LO and the NLO EW approximations. 
Lower panel: relative difference, in percent, between the 
full QCD$\otimes$EW predictions and the pure QCD ones (green, dashed line), in comparison 
with the relative contribution due to pure NLO EW corrections (red, solid line).}
\label{combination1} 
\end{figure}

\begin{figure} 
\begin{center}
\includegraphics[height=6.cm]{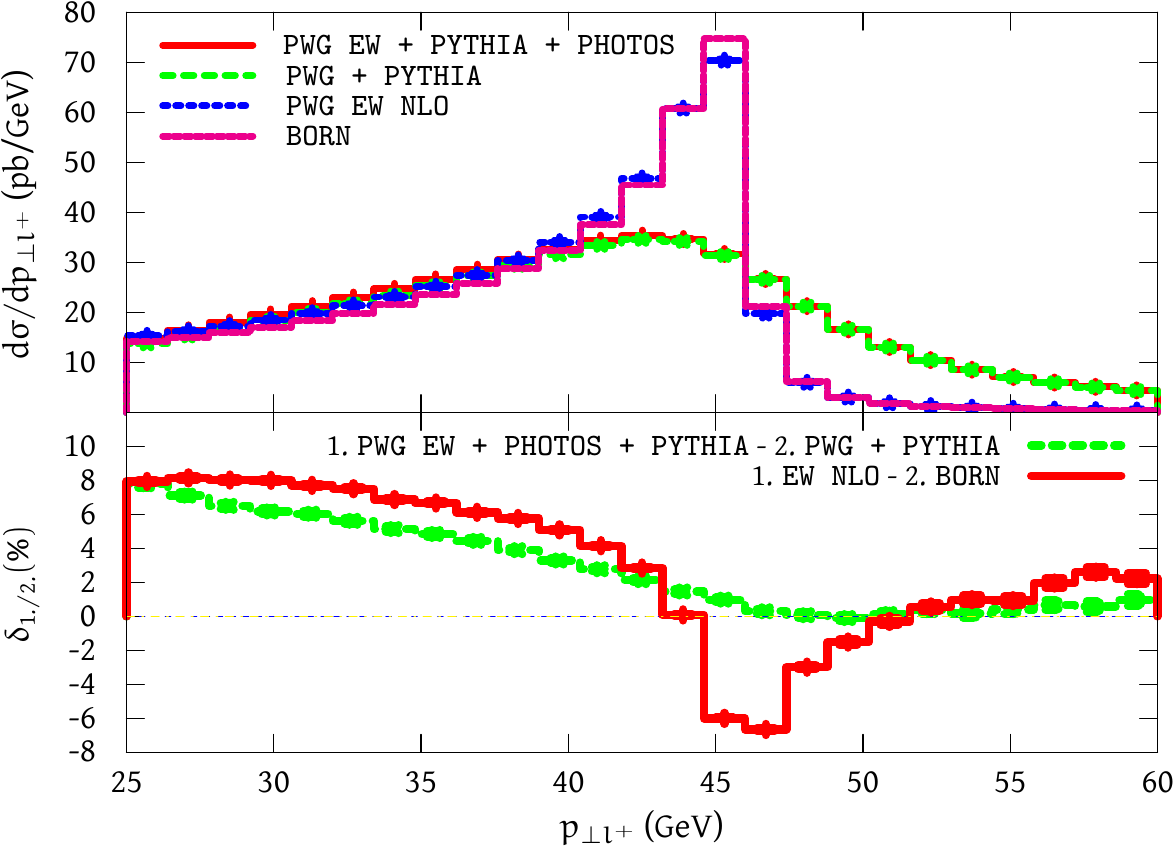}
\end{center}
\caption{The same as Fig. \ref{combination1} for the lepton transverse momentum distribution.}
\label{combination2}
\end{figure}

\begin{figure} 
\begin{center}
\includegraphics[height=6.cm]{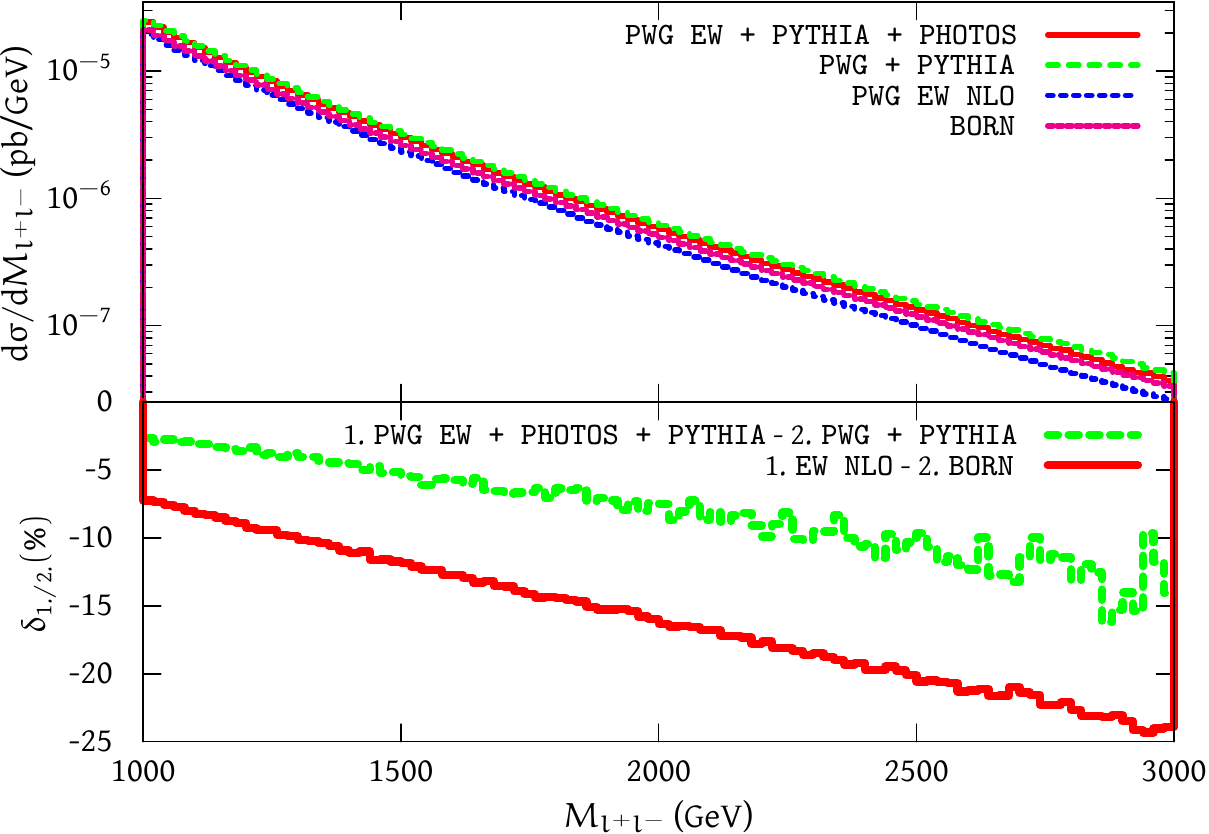}
\end{center}
\caption{The same as Fig. \ref{combination1} in the region $M_{l^+ l^-}$ > 1~TeV.}
\label{combination3}
\end{figure}

\begin{figure} 
\begin{center}
\includegraphics[height=6.cm]{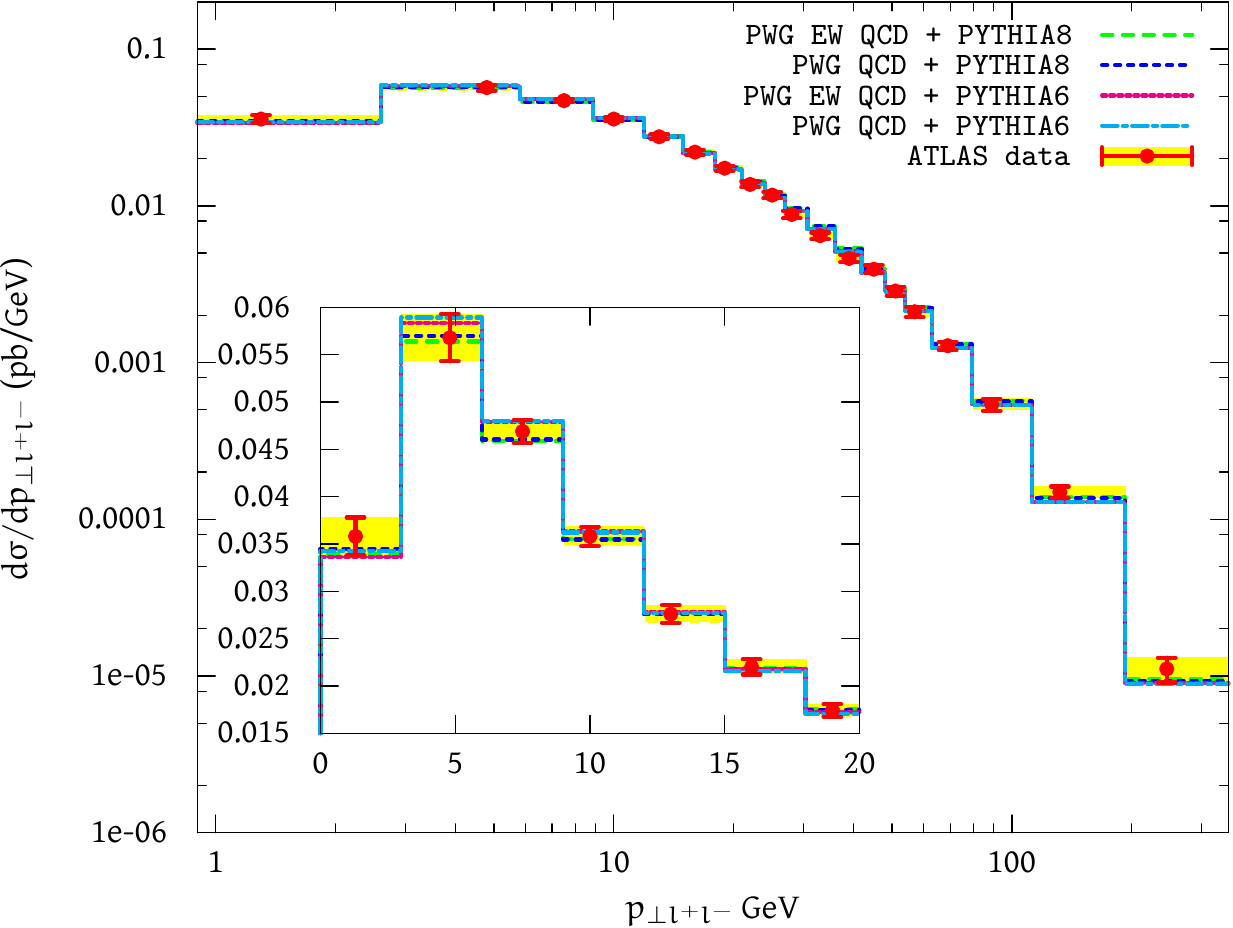}
\end{center}
\caption{The normalized differential cross section as a function of $p_\perp^Z$ for the full range 
up to 350 GeV (whole plot) and 
in the range $p_\perp^Z$ < 20~GeV (inset). The ATLAS data (for bare muons) are compared with 
the predictions of the POWHEG BOX with full QCD$\otimes$EW contributions and QCD corrections only, and 
according to two PYTHIA versions.}
\label{combination4}
\end{figure}

\begin{figure} 
\begin{center}
\includegraphics[height=6.cm]{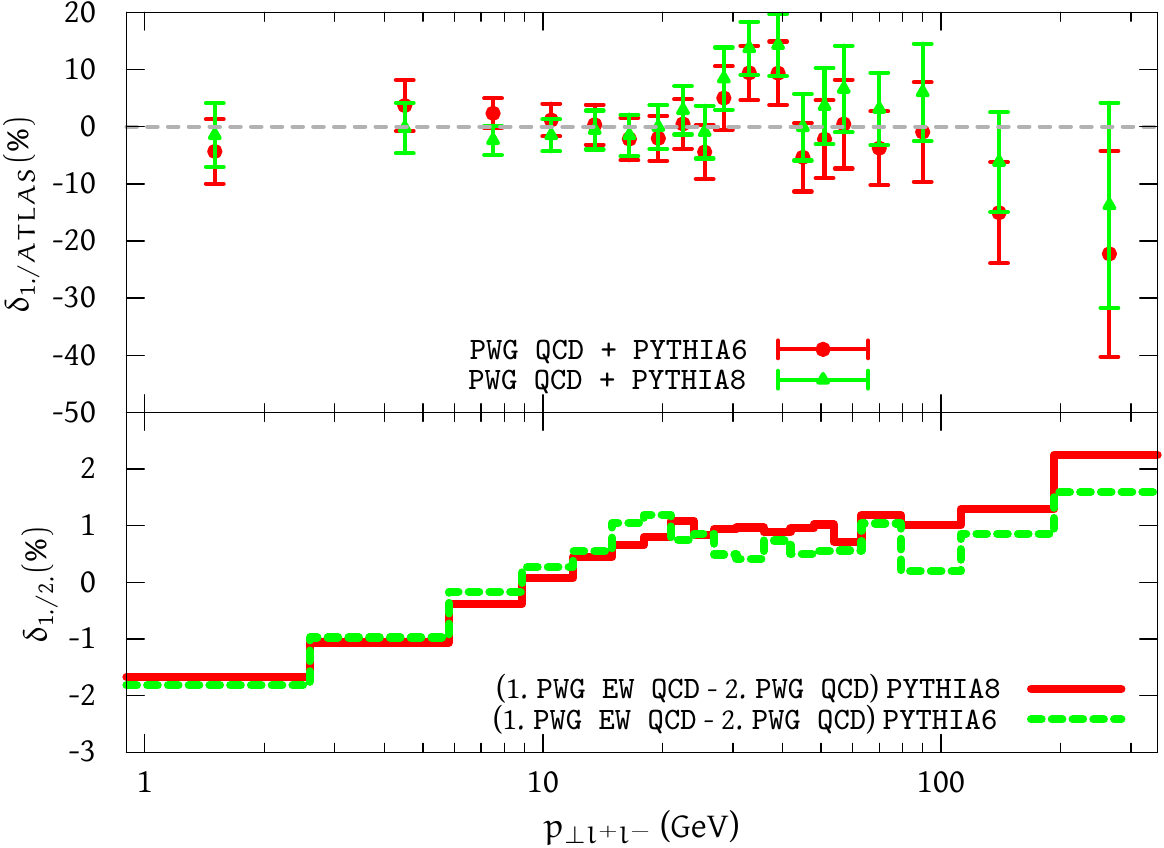}
\end{center}
\caption{Upper panel: ratio of the POWHEG BOX QCD predictions over the ATLAS data, according to two 
different version of PYTHIA. Lower panel: 
relative difference, in percent, between the 
full QCD$\otimes$EW predictions and the pure QCD ones of POWHEG BOX for two different PYTHIA versions.}
\label{combination5}
\end{figure}

\begin{figure} 
\begin{center}
\includegraphics[height=6.cm]{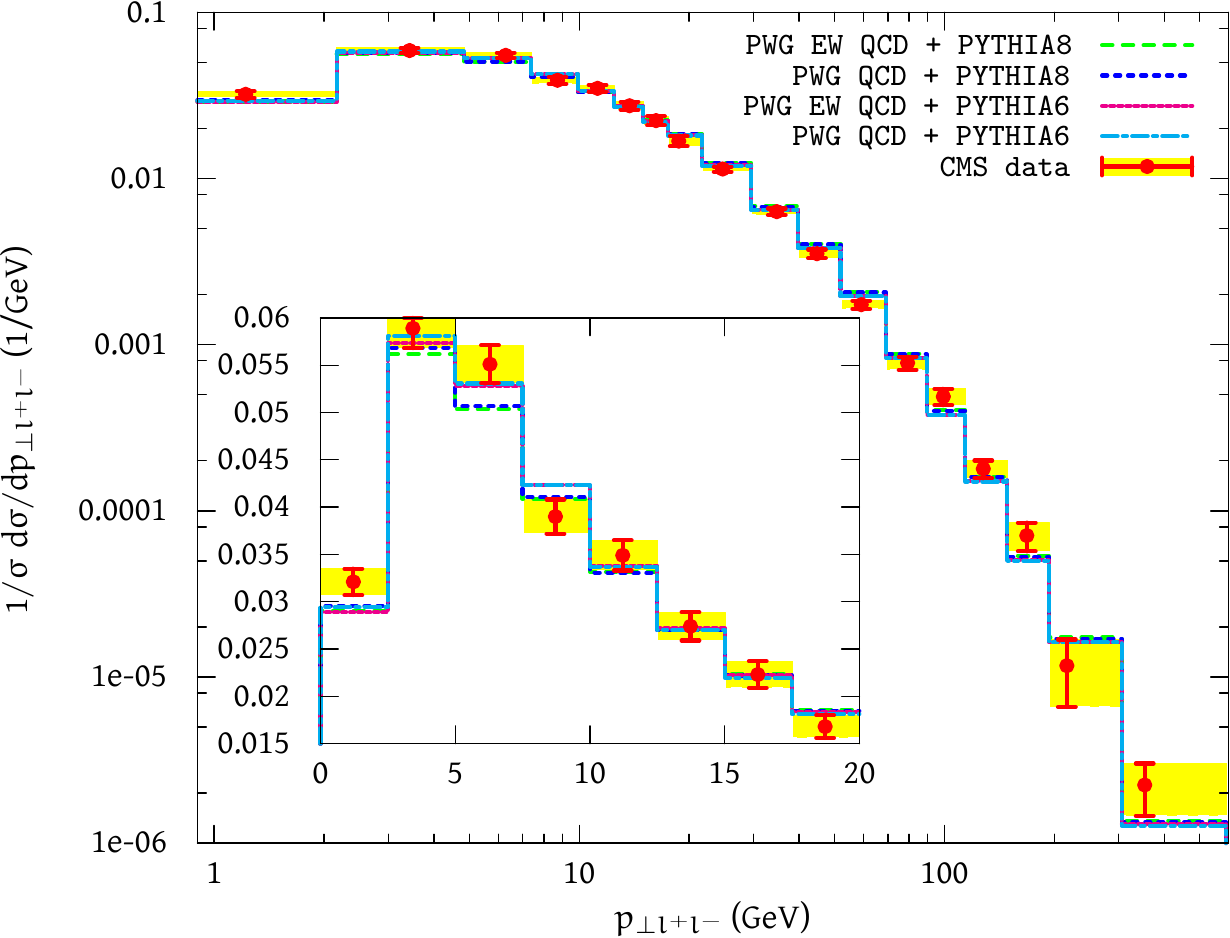}
\end{center}
\caption{The same as Fig. \ref{combination4} for CMS data (for bare muons) up to 600~GeV.}
\label{combination6}
\end{figure}

\begin{figure} 
\begin{center}
\includegraphics[height=6.cm]{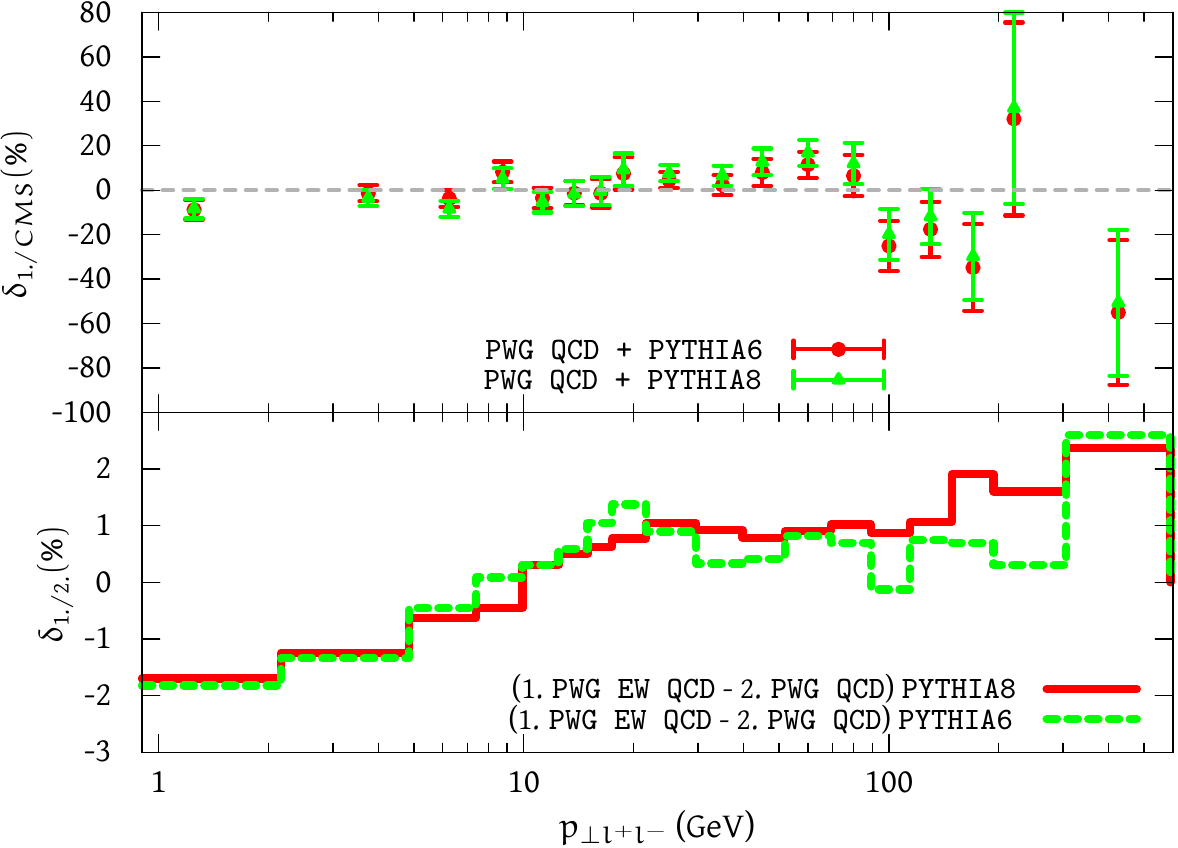}
\end{center}
\caption{The same as Fig. \ref{combination5} for CMS data.}
\label{combination7}
\end{figure}

The full results of the POWHEG BOX for the combined effect of QCD and EW 
corrections to the invariant mass and lepton transverse momentum distribution are 
shown in Fig.~\ref{combination1} - Fig.~\ref{combination2}
in the resonance region, and in Fig.~\ref{combination3}
for $M_{l^+ l^-}$ > 1~TeV. Further numerical results concerning the 
transverse momentum distribution of the $Z$ boson in the resonance region are shown in 
Fig.~\ref{combination4} - Fig.~\ref{combination7}, in comparison with 
ATLAS~\cite{Aad:2011gj} and CMS~\cite{Chatrchyan:2011wt}  data at $\sqrt{s} = 7$~TeV.

For Fig.~\ref{combination1} - Fig.~\ref{combination3} 
the complete predictions have been obtained by matching
the NLO QCD and EW corrections with QCD (PYTHIA version 6.4 \cite{Sjostrand:2006za}) and 
QED (PHOTOS) showers.
For the sake of comparison, the pure QCD predictions of the standard POWHEG BOX are
also given, together with the pure NLO EW results. The absolute predictions for the
various distributions are shown in the upper panels of each plot. The lower panels display
the relative difference, in percent, between the results of the new version of the POWHEG
BOX and the standard QCD release, as well as the relative effect due to the genuine NLO EW
corrections. Therefore the comparison between the two lines in each lower panel 
of  Fig.~\ref{combination1} - Fig.~\ref{combination3} provides
a measure of the QCD$\otimes$EW factorization and, more precisely, of mixed 
leading logarithmic corrections at the order $\aem^m \alpha_s^n, m,n \geq 1$.

Generally speaking, one can notice that both QCD and EW corrections (and their combination) are 
necessary for a proper control of the normalization and shape of the distributions.
Particularly, the invariant mass distribution in the resonance region receives large 
corrections and a significative shape modification from QED corrections, as 
emphasized in previous studies \cite{CarloniCalame:2007cd,Baur:1997wa,CarloniCalame:2005vc}. 
As shown in Fig. \ref{combination1}, the left tail of the distribution is enhanced by the mechanism 
of final state photon radiation by several tens of percent, while the peak value is reduced by about 20\%. 
For this distribution the effect of the EW corrections largely exceeds that of QCD radiation, and the 
impact of mixed QED$\otimes$EW contributions is substantial, especially in the left tail of the 
distribution. The same kind of effect is present in the forward-backward asymmetry as a function 
of the lepton pair invariant mass for $M_{l^+ l^-}$ below the $Z$ mass, as we checked explicitly 
in our simulations. For the lepton $p_\perp$ 
the well-known overwhelming QCD effects are by far dominant over the 
EW contributions, whose shape is washed out  by QCD radiation (see Fig.~\ref{combination2}). 
For such a distribution, the interplay of the particularly large QCD corrections with the 
ten percent level EW effects gives rise to mixed contributions of the order of
several percents close to the peak, as clearly visible in the lower panel of Fig.~\ref{combination2}. 

Non-negligible QCD$\otimes$EW corrections are also present in the very high tail of 
the invariant mass distribution shown in Fig. \ref{combination3}. 
In this region the large EW corrections, enhanced by Sudakov-like logarithms, in association with QCD radiation
induce mixed contributions which grow from a few to several percents, as can be 
seen in the lower panel of Fig.~\ref{combination3}. 

Last but not least, we show for illustrative purposes in Fig.~\ref{combination4} - 
Fig.~\ref{combination7} our results for the $Z$ transverse momentum 
distribution, whose knowledge is crucial for a precise measurement of the $W$ mass and whose description 
represents a challenge for QCD. The predictions of the POWHEG BOX are compared with 
ATLAS data \cite{Aad:2011gj} (see 
Fig.~\ref{combination4} - Fig.~\ref{combination5})
and CMS data \cite{Chatrchyan:2011wt}  (see Fig.~\ref{combination6} - Fig.~\ref{combination7}) 
at $\sqrt{s}$ = 7 TeV\footnote{We used the data available at the HepData repository http://hepdata.cedar.ac.uk.}. 
For these results, we imposed the acceptance and invariant mass cuts quoted in the experimental papers, we 
focused on the data referring to bare muons and we 
used the MSTW2008 NLO PDF set~\cite{Martin:2009iq}. Because of the considerable dependence of the $p_\perp^Z$ 
distribution on adjustable non-perturbative parameters, we considered POWHEG interfaced to two different versions 
of PYTHIA generator, na\-me\-ly PYTHIA version 6.4 \cite{Sjostrand:2006za} and version 8.1~\cite{Sjostrand:2007gs}. 
While higher-order QED radiation is treated with 
PHOTOS when POWHEG is interfaced to PYTHIA~6.4, the simulation performed with 
PYTHIA 8.1 uses its internal available tool for QED shower.  
We included the contribution of hadronization in both PYTHIA versions and 
changed the  PYTHIA 8.1 default for the treatment of the QED Shower $\aem (p_\perp)$ 
with $\aem (0)$, for consistency with PHOTOS.

In Fig. \ref{combination4} and Fig. \ref{combination6} we show the absolute predictions of POWHEG for the 
normalized differential cross section as a function of $p_\perp^Z$ in comparison with ATLAS and CMS 
data, respectively. The POWHEG results are presented both with and 
without NLO EW and QED Shower corrections, 
and according to the two aforementioned PYTHIA versions. The inset figures show the data-theory comparison in 
the low $p_\perp^Z$ region. Albeit an optimal description of the $Z$ transverse momentum 
distribution over the full range seems
to be problematic, there is a rather good agreement with the data, especially when 
considering moderate $p_\perp^Z$ values, as can be appreciated by looking 
at the upper panels of 
Fig. \ref{combination5} and Fig. \ref{combination7}.
They show the ratio 
of the POWHEG BOX QCD predictions over the ATLAS and CMS data, as obtained with PYTHIA 
version 6.4 and version 8.1, respectively. 
The lower panels of Fig. \ref{combination5} and Fig. \ref{combination7} show in conclusion the 
impact on the $Z$ transverse momentum distribution due to NLO EW and multiple photon corrections, as 
well as to $O(\aem \alpha_s)$ contributions. As can be noticed, these contributions are largely 
independent of the particular version of the 
QCD/QED Shower generator under consideration. They introduce a 
correction of a few percents in the whole $p_\perp^Z$ range.
 
As a concluding remark, it is worth noting the the possibility of
simulating the contribution of mixed QCD$\otimes$EW corrections with the new
tool is due to the particular factorized form of the EW corrections in
the QCD POWHEG framework. In this respect, our predictions contain 
complementary  information  with respect to the calculation of 
Refs. \cite{Bardin:2012jk,Bondarenko:2013nu,Dittmaier:2009cr,Li:2012wn}, where QCD and EW 
corrections are combined additively and QCD PS or multi-photon radiation contributions 
are not taken into account. In those calculations, mixed QCD-EW corrections are neglected, by construction, 
whi\-le they are included, together with
higher-order QCD and QED contributions, in our formulation, 
as detailed in Section~\ref{section:calculation}. 
On the other hand, the main limitation and source of theoretical error in 
our predictions comes from the NNLO QCD corrections available in 
the additive combination of Ref.~ \cite{Li:2012wn} but missing in our calculation. 
They are known
\cite{Catani:2009sm,Melnikov:2006kv,Gavin:2010az} to change the NLO QCD 
normalization of some per cents for sufficiently inclusive differential cross sections in the 
presence of standard selection cuts.  Moreover, the NNLO QCD corrections
provide, through $Z$ + jet production at NLO accuracy, a relevant 
contribution to the observables particularly sensitive to 
hard QCD radiation, like the $Z$ transverse momentum at high $p_T$, 
where our calculation is limited to a LO accuracy.


\section{Conclusions}
\label{section:conclusions}
We have generalized previous work on the inclusion of the EW corrections to 
single $W$ production in the POWHEG BOX
to cope with the NC DY process. We have added NLO EW and QED multiple photon corrections to the native 
NLO and PS QCD contributions. To this aim, we have exploited the 
process-independent structure of the POWHEG BOX
framework and resorted to various general techniques already successfully developed for $W$-boson production.

We have provided evidence of the accuracy of the approach and presented a sample of phenomenological results 
about the combination of QCD and EW corrections to lepton pair production in hadronic collisions at LHC energies. 
In particular, we have shown that the leading contribution of mixed QCD$\otimes$EW 
 corrections may affect some observables at the level of several percents, 
beyond the separate effect of strong, weak and electromagnetic corrections. Therefore, the
exact calculation of $O(\aem \alpha_s)$ corrections would be desirable for a better 
theoretical control of the DY cross sections.

The new tool enables to obtain precise and realistic predictions for the NC DY process, in the same way as the
charged-current channel. It is available at the web site of  POWHEG BOX. Further theoretical ingredients, such as
pho\-ton-induced processes~\cite{CarloniCalame:2007cd,Dittmaier:2009cr,Arbuzov:2007kp,Brensing:2007qm}
and presently neglected higher-order corrections, will be made available in future releases. 

We expect that the results here presented will facilitate the work of experimental analysis and 
data interpretation at 
hadron colliders and will allow in the future a rather simple inclusion in the POWHEG BOX of EW effects 
to further processes of physics interest.

\begin{acknowledgements}
We are grateful to various colleagues of the LHC and Tevatron communities for interest 
in our work and many useful discussions.
This work was supported in part by the Research Executive Agency (REA) of the 
European Union under the Grant Agreement number 
PITN-GA-2010-264564 (LHCPhenoNet), and by the Italian Ministry of University and Research under the 
PRIN program 2010-2011. The work of L.B. is supported by the 
ERC grant 291377, ``LHCtheory - 
Theoretical predictions and analyses of LHC physics: advancing the precision frontier".
F.P. would like to thank the CERN PH-TH Department for partial support and 
hospitality during several stages of the work.
\end{acknowledgements}

\end{document}